\documentclass[a4paper,11pt]{article}
\pdfoutput=1 
\usepackage{jheppub} 

\usepackage[T1]{fontenc} 
\usepackage{caption} 
\usepackage{subcaption}

\usepackage[latin9]{inputenc}
\usepackage{float}
\usepackage{amsmath}
\usepackage{amssymb}
\usepackage{graphicx}
\usepackage{esint}
 \usepackage{hyperref}
\usepackage{comment}
\usepackage{color}
\usepackage{microtype}
\usepackage{cleveref}
\usepackage{breakurl}
\usepackage{bbm}
\newcommand{\be}{\begin{equation}}
\newcommand{\ee}{\end{equation}}
\newcommand{\ben}{\begin{displaymath}}
\newcommand{\een}{\end{displaymath}}
\newcommand{\bea}{\begin{eqnarray}}
\newcommand{\eea}{\end{eqnarray}}

   \newcommand{\rf}[1]{(\ref{#1})}
\newcommand{\vp}{\varphi}

\def\be{\begin{equation}}
\def\ee{\end{equation}}
\def\bea{\begin{eqnarray}}
\def\eea{\end{eqnarray}}
\def\ba{\begin{array}}
\def\ea{\end{array}}
\def\bit{\begin{itemize}}
\def\eit{\end{itemize}}

\def\vp{\varphi}

 \makeatletter
 \usepackage{hyperref}

\allowdisplaybreaks

\makeatother

\makeatletter
\DeclareRobustCommand{\rcite}[1]{%
  \rcite@aux#1,\@nil{#1}%
}
\def\rcite@aux#1,#2\@nil#3{%
  \if\relax#2\relax
    Ref.~\cite{#3}%
  \else
    Refs.~\cite{#3}%
  \fi
}
\makeatother

\hypersetup{
    colorlinks = true,
    citecolor = {blue},
    linkcolor = {blue},
    urlcolor = {blue},
}

 \title{\rm {\bf \huge  \boldmath On hilltop and brane inflation after Planck}}

\author{Renata Kallosh}
\author{and  Andrei Linde}

\affiliation{Stanford Institute for Theoretical Physics and Department of Physics, Stanford University, Stanford, CA 94305, USA}
\emailAdd{kallosh@stanford.edu}
 \emailAdd{alinde@stanford.edu}

\notoc

\abstract{Hilltop inflation models are often described by potentials $V = V_{0}(1-{\phi^{n}\over m^{{n}}}+...)$. The omitted terms indicated by ellipsis do not affect inflation  for $m \lesssim 1$, but the most popular models with $n =2$ and $4$ for $m \lesssim 1$ are 
ruled out observationally. Meanwhile  in the large $m$ limit the results of the calculations of the tensor to scalar ratio $r$ in the models with $V = V_{0}(1-{\phi^{n}\over m^{{n}}})$, for all $n$, converge to $r= 4/N \lesssim 0.07$, as in chaotic inflation with $V \sim \phi$,  suggesting a reasonably good fit to the Planck data. We show, however, that this  is an artifact  related to the inconsistency of the model $V = V_{0}(1-{\phi^{n}\over m^{{n}}})$ at $\phi > m$. Consistent generalizations of this model in the large $m$ limit typically lead to a much greater value  $r= 8/N$,  which negatively affects the observational status of hilltop inflation.  Similar results are valid for D-brane inflation with $V = V_{0}(1-{m^{n}\over \phi^{{n}}})$, but consistent generalizations of D-brane inflation models may successfully complement  $\alpha$-attractors in describing most of the area in the ($n_{s}$, $r$) space favored by Planck 2018.}

\begin{document}

\maketitle

 \newpage 

\vskip 1 cm

 \parskip 3pt
\section{Introduction}

The Planck 2018 data release  \cite{Akrami:2018odb}  provides a detailed review of a large family of inflationary models in comparison with observational data. Some of the most important results are described in Sect. 4.2 of  \cite{Akrami:2018odb}, especially in the Table 5, and also in the Fig. 8, which we reproduce below.  
 \begin{figure}[h!]
\begin{center}
\includegraphics[scale=1.1]{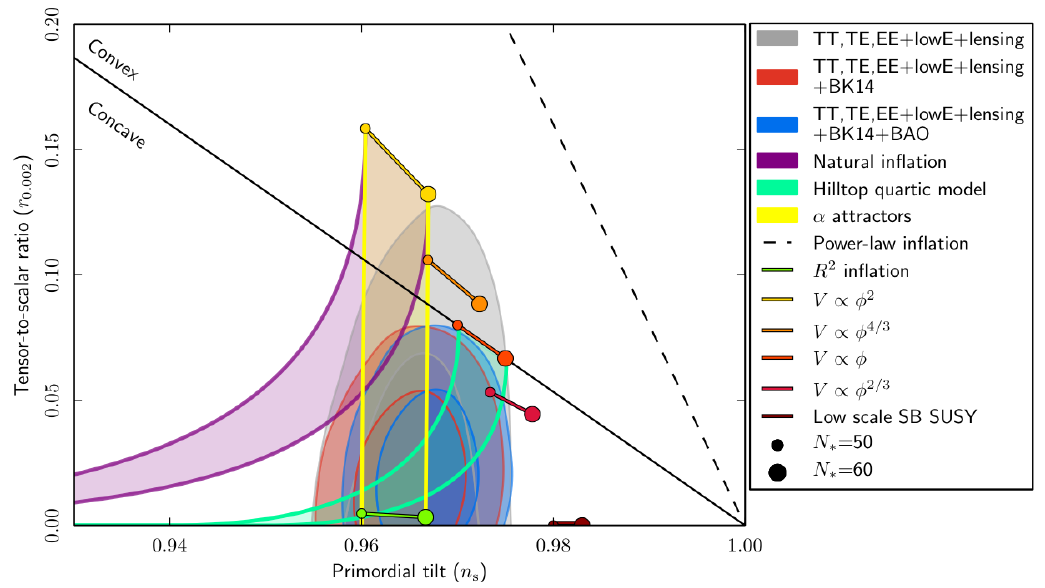}
\end{center}
\vskip -0.5cm 
\caption{\footnotesize Inflationary models and the  Planck 2018 results, according to  \cite{Akrami:2018odb}.}
\label{F0}
\end{figure} 
The models providing the best fit to the observational data include various versions of $\alpha$-attractors shown at Fig. \ref{F0} by a vertical yellow stripe, as well as the Starobinsky model and the Higgs inflation model  shown by the small green disks at the lower end of the yellow  $\alpha$-attractor stripe. Another set of models matching the Planck data are the hilltop models with the potentials $V = V_{0}(1-{\phi^{n}\over m^{{n}}}+...)$ and the so-called D-brane inflation models with the potentials $V = V_{0}(1-{m^{n}\over \phi^{{n}}}+...)$  for $n = 2$ and $n = 4$, see Table 5 in  \cite{Akrami:2018odb}.   In the Planck 2013 and 2015 data releases  \cite{Planck:2013jfk,Ade:2015lrj} it was emphasized that the omitted higher order terms shown by ellipsis are supposed to stabilize the potential from below.  Moreover, it was pointed out in   \cite{Ade:2015lrj} that consistent modifications of the hilltop model with $n = 2$, such as the simplest Higgs model $V = V_{0}(1-{\phi^{2}\over m^{{2}}})^{2}$, do not fit the data too well. The subsequent numerical analysis  was performed for the models  $V = V_{0}(1-{\phi^{n}\over m^{{n}}})$, with emphasis on $n = 4$.  The results  for $n = 4$  are shown by the green area in Fig. \ref{F0}. Similarly, the analysis of  D-brane inflation was performed for  $V = V_{0}(1-{m^{n}\over \phi^{{n}}})$, see  Table 5 in \cite{Akrami:2018odb}.

These results play important role not only in evaluation of the presently available inflationary models, but also in the planning of the new generations of the cosmological observations. In particular, Fig.~10 in the CMB-S4 Science Book \cite{Abazajian:2016yjj}, Fig. 1 of the CMB-S4 Science Case \cite{Abazajian:2019eic}, and Fig.~2.2 in the PICO report  \cite{Hanany:2019lle} illustrate the present status of inflationary models related to their plans by  the vertical stripe describing $\alpha$-attractors, as well as the green area corresponding to the hilltop models  with the potentials $V = V_{0}(1-{\phi^{4}\over m^{{4}}})$. That is why we decided to revisit these three classes of models favored by the Planck data. Since we extensively studied $\alpha$-attractors in many other papers, we will only briefly review them in this paper, and concentrate on the hilltop and D-brane inflation models.

A large family of cosmological attractors, including the $\alpha$-attractors, was discovered shortly after the Planck 2013 data release \cite{Kallosh:2013hoa,Kallosh:2013tua,Ferrara:2013rsa,Kallosh:2013yoa,Kallosh:2014rga,Kallosh:2014laa,Galante:2014ifa}. It was a result of our attempts to understand the mysterious nearly exact coincidence of the cosmological predictions of the Starobinsky model \cite{Starobinsky:1980te} and the Higgs inflation model \cite{Salopek:1988qh,Bezrukov:2007ep}. We found that the main reason of this coincidence is that at the intermediate steps of the transformation from the original formulation of these models to the Einstein frame one encounters a scalar field with a kinetic term having a quadratic pole at a certain  value of the field. This singularity disappears after transition to canonical variables. 
 
A similar pole appears also in the model of conformal attractors \cite{Kallosh:2013hoa}, which lead to the same predictions $n_{s} =1-2/N$ and $r=12/N$ as the Starobinsky model and the Higgs inflation model. Here $N$ is the number of e-foldings. An advantage of conformal attractors is that these predictions are practically independent on the choice of the original potential, which is why we called these models ``attractors''. This property is shared by the generalization of the Higgs inflation proposed in  \cite{Kallosh:2013tua}. Finally, a more general family of cosmological attractors was found, called $\alpha$-attractors \cite{Ferrara:2013rsa,Kallosh:2013yoa,Kallosh:2014rga,Kallosh:2014laa,Galante:2014ifa}. In the  supergravity generalizations of this model, the pole appears as a result of the underlying hyperbolic geometry of the moduli space \cite{Carrasco:2015uma}.  In these models one has
\be\label{alphansr}
n_{s} =1-{2\over N}, \qquad r={12\alpha \over N} \ , 
\ee
which coincides with the predictions of the Starobinsky model, Higgs inflation and conformal attractors for $\alpha = 1$, and with the predictions of the GL model  \cite{Goncharov:1985yu,Linde:2014hfa} for $\alpha = 1/9$. 
The $\phi^2$ potential corresponds to the  limit of infinite 
 $\alpha$ and this parameter  decreases with decreasing $r$. With  $N$  in the interval from $50$ to $60$,  these predictions span the range shown by the yellow area in   Fig. \ref{F0}.

In this paper we will briefly describe $\alpha$-attractors, and then turn our attention to hilltop inflation models with $V = V_{0}(1-{\phi^{n}\over m^{{n}}}+...)$ \cite{Linde:1981mu,Kinney:1995cc,Boubekeur:2005zm} and  D-brane inflation with 
$V = V_{0}(1-{m^{n}\over \phi^{{n}}}+...)$ \cite{Dvali:2001fw,Burgess:2001fx,GarciaBellido:2001ky,Kachru:2003sx,Martin:2013tda,Kallosh:2018zsi}. In the limit $m\ll 1$, the predictions of both models are fairly stable with respect to the unspecified higher order terms shown by ellipsis in these equations. The most interesting versions of both models correspond to $n= 4$. The hilltop inflation with $n= 4$ in the limit $m\ll 1$  describes an attractor with 
\be\label{hill4} 
n_{s} =1-{3\over N}, \qquad r={8 m^{4}  \over N^{3}} \ .
\ee  
For $N = 50$ one has $n_{s}  = 0.94$, and for $N = 60$ one has $n_{s} = 0.95$. Both results  strongly disagree   with observational data. 

However, if one considers an opposite limit $m \gg 1$ and ignores the ellipsis, reducing the theory to  $V = V_{0}(1-{\phi^{n}\over m^{{n}}})$, then one finds that in the large $m$ limit the predictions of this theory converge to the predictions of inflation with a linear potential $V \sim \phi$,
\be\label{lin} 
n_{s} = 1-{3\over 2N}, \qquad r={4 \over N} \ .
\ee
The green area in Fig. \ref{F0} describing this theory interpolates between the two attractor points \rf{hill4} and \rf{lin}. Both attractor points are bad: $n_{s} =1-{3\over N}$ in \rf{hill4}   is too small, and $r=4/N$ in \rf{lin} is too large, but  on its way between these two  attractors, the green area significantly overlaps with the area favored by Planck 2018. 

The situation becomes even more interesting if one takes into account that in the large $m$ limit the predictions of the model with a linear potential \rf{lin} is an attractor point for the hilltop inflation with $V = V_{0}(1-{\phi^{n}\over m^{{n}}})$ not only for $n = 4$, but  for any $n$, and not only for the hilltop inflation, but also for the D-brane inflation with  $V = V_{0}(1-{m^{n}\over \phi^{{n}}})$  \cite{Martin:2013tda}. 

One may wonder why such a large class of different rather complicated models leads to the same prediction as the simplest model with the linear potential?  Could it be that by revealing the deep physical reason of this remarkable result one can learn something important, just as we did when we tried to understand  the nearly exact coincidence of the results of the Starobinsky and the Higgs inflation model?

Unfortunately, as we will see, this mysterious result is not so mysterious after all. The  common feature of hilltop models $V = V_{0}(1-{\phi^{n}\over m^{{n}}})$ and D-brane inflation models $V = V_{0}(1-{m^{n}\over \phi^{{n}}})$ is that in the large $m$ limit the last 60 e-foldings of inflation in these models occur when the field $\phi$  moves from $\phi \sim m-10$ to $m-1/\sqrt 2$ (from $\phi \sim m+10$ to $m+1/\sqrt 2$ for the D-brane models). For $m \gg 10$ and $\phi> 0$, these potentials look like a straight line $V \propto  m-\phi$ ($V \propto  \phi-m$ for D-brane models), which explains why all of these potentials have the same prediction \rf{lin} in the large $m$ limit. 

But this means that the remarkable coincidence discussed above is directly related to the fact that the end of inflation in these models occurs shortly before the inflaton field hits the point $\phi = m$ with $V(\phi) = 0$, falls towards $V(\phi) < 0$, and the whole universe collapses. Thus such models are not good. 

The models with the potentials unbounded from below are bad for any $m$. Therefore in the description of these models in the Planck 2013 data release it was emphasized that the higher order terms should stabilize the vacuum. For $m\lesssim 1$, one can easily add  terms  higher order in ${\phi\over m}$, forming a minimum of the potential at $V \geq 0$, without altering inflationary predictions of the original model. 
But the most natural attempts to improve these models at $m \gg1$ significantly change their predictions. If we want to stick to the original design, considering the potentials $V({\phi\over m})$, then  in the large $m$ limit the modified hilltop potentials near their minima in the first approximation are not linear but quadratic, on  scale $\Delta \phi \equiv |m-\phi|  \gg 1$. This is a rather general result described in \cite{Kallosh:2014rga,Kallosh:2014laa}  in a different context.  Therefore in the large $m$  limit the predictions of these models  converge to the predictions of the chaotic inflation potential $V \sim \phi^{2}$, which yields 
\be\label{quadr} 
n_{s} = 1-{2\over N}, \qquad r={8\over N} \ .
\ee
 In application to consistent versions of the quadratic  hilltop models $V = V_{0}(1-{\phi^{2}\over m^{{2}}})$, such as the Higgs potential $V = V_{0}(1-{\phi^{2}\over m^{{2}}})^{2}$, this result was already discussed in the Planck 2015 data release \cite{Ade:2015lrj}, but we found that it is valid for any $n$, including $n = 4$.  Thus the  improved versions of the hilltop models in the large $m$ limit typically predict $r={8\over N}$, which is {\it twice as large} as the naive prediction $r={4\over N}$ of the  hilltop models with the potentials $V = V_{0}(1-{\phi^{n}\over m^{{n}}})$.

A similar problem affects the D-brane inflation models $V = V_{0}(1-{m^{n}\over \phi^{{n}}})$, where the improved versions of these models typically lead to a much greater value of $r$ in the large $m$ limit. Now that we are aware of this problem, we will take a second look at all of these models, starting with $\alpha$-attractors, and then discuss hilltop and D-brane models in detail, going one step beyond the simplest versions of these theories.

\section{\boldmath $\alpha$-attractors}

The basic principles of the theory of $\alpha$-attractors can be explained using a simple toy model with the Lagrangian  \cite{Galante:2014ifa,Kallosh:2015zsa}
 \be
 {1\over \sqrt{-g}} \mathcal{L} = { R\over 2}   -  {(\partial_{\mu} \phi)^2\over 2(1-{\phi^{2}\over 6\alpha})^{2}} - V(\phi)   \,  .
\label{cosmo}\ee
Here $\phi(x)$ is the scalar field, the inflaton.  The origin of the pole in the kinetic term can be explained in the context of hyperbolic geometry in supergravity and string theory.
The parameter  $\alpha$  can take any positive value.  Instead of the variable $\phi$, one can use a canonically normalized field $\vp$ by solving the equation ${\partial \phi\over 1-{\phi^{2}\over 6\alpha}} = \partial\vp$, which yields $
\phi = \sqrt {6 \alpha}\, \tanh{\varphi\over\sqrt {6 \alpha}}$.
The full theory, in terms of the canonical variables, becomes
 \be
 {1\over \sqrt{-g}} \mathcal{L} = { R\over 2}   -  {(\partial_{\mu}\varphi)^{2} \over 2}  - V\big(\sqrt {6 \alpha}\, \tanh{\varphi\over\sqrt {6 \alpha}}\big)   \,  .
\label{cosmoqq}\ee
 If  the potential and its derivatives are non-singular at $\phi =  \sqrt {6 \alpha}$, the asymptotic behavior of the potential at large $\varphi>0$ is given by
\be\label{plateau}
V(\vp) = V_{0} - 2  \sqrt{6\alpha}\,V'_{0} \ e^{-\sqrt{2\over 3\alpha} \varphi } \ .
\ee
Here  $V_0 = V(\phi)|_{\phi =  \sqrt {6 \alpha}}$ is the height of the plateau potential, and $V'_{0} = \partial_{\phi}V |_{\phi = \sqrt {6 \alpha}}$. Importantly, the coefficient $2  \sqrt{6\alpha}\,V'_{0}$ in front of the exponent  can be absorbed into a redefinition (shift) of the field $\varphi$. Therefore all inflationary predictions of this theory in the regime with $e^{-\sqrt{2\over 3\alpha} \varphi } \ll 1$ are determined only by two parameters, $V_{0}$ and $\alpha$.

The amplitude of inflationary perturbations $A_{s}$ in these models matches the Planck normalization for  $ {V_{0}\over  \alpha} \sim 10^{{-10}}$.  For the simplest  model $V = {m^{2}\over 2} \phi^{2}$, belonging to a class of T-models  with the potential symmetric with respect to $\phi \to -\phi$, one finds
\be\label{T}
V =  3m^{2 }\alpha \tanh^{2}{\varphi\over\sqrt {6 \alpha}} \ .
\ee
Then  the condition $ {V_{0}\over  \alpha} \sim 10^{{-10}}$ reads $ m  \sim   0.6 \times10^{{-5}}$. The cosmological predictions of this model  are shown by the  yellow vertical stripe in Fig. \ref{F0}. 

We should note that even though the predictions for large $N$ and small $\alpha$ are rather well defined, the value of $N$ itself does depend on the mechanism of reheating and post-inflationary equation of state, which is reflected in the uncertainty of the choice between $N\sim 50$ and $N \sim 60$. Also, predictions of different versions of $\alpha$-attractors converge to their target \rf{alphansr} in a slightly different way.   
For example,   one may start with a model 
\be
 {1\over \sqrt{-g}} \mathcal{L} = { R\over 2}   -  {3\alpha\over 4}  \left({\partial_{\mu} t  \over t}\right)^{2} - V(t)   \,  .
\label{Eorig}
\ee
One can represent this model in terms of  canonical variables  $\varphi$, where $t = e^{-\sqrt{2\over 3\alpha} \varphi}$. For $V= V_{0}(1-t)^{2} $ this leads to a class of E-models with
\be\label{E}
V(\varphi) = V_{0}\left(1-e^{-\sqrt{2\over 3\alpha} \varphi}\right)^{2}, 
\ee
which coincides with the potential in the Starobinsky model for $\alpha = 1$. The difference between this  model for $\alpha = 1$ and the original Starobinsky model \cite{Starobinsky:1980te} is that adding extra terms $R^{n}$ to the Starobinsky model \cite{Starobinsky:1980te} can easily affect the plateau shape of its potential, whereas the asymptotic plateau shape of the $\alpha$-attractor potential \rf{E} at large $\varphi$ is quite stable with respect to  considerable modifications of the original potential $V(t)$, and, as a consequence, stable  with respect to quantum corrections \cite{Kallosh:2016gqp}.

 \begin{figure}[t!]
\begin{center}
\includegraphics[scale=0.45]{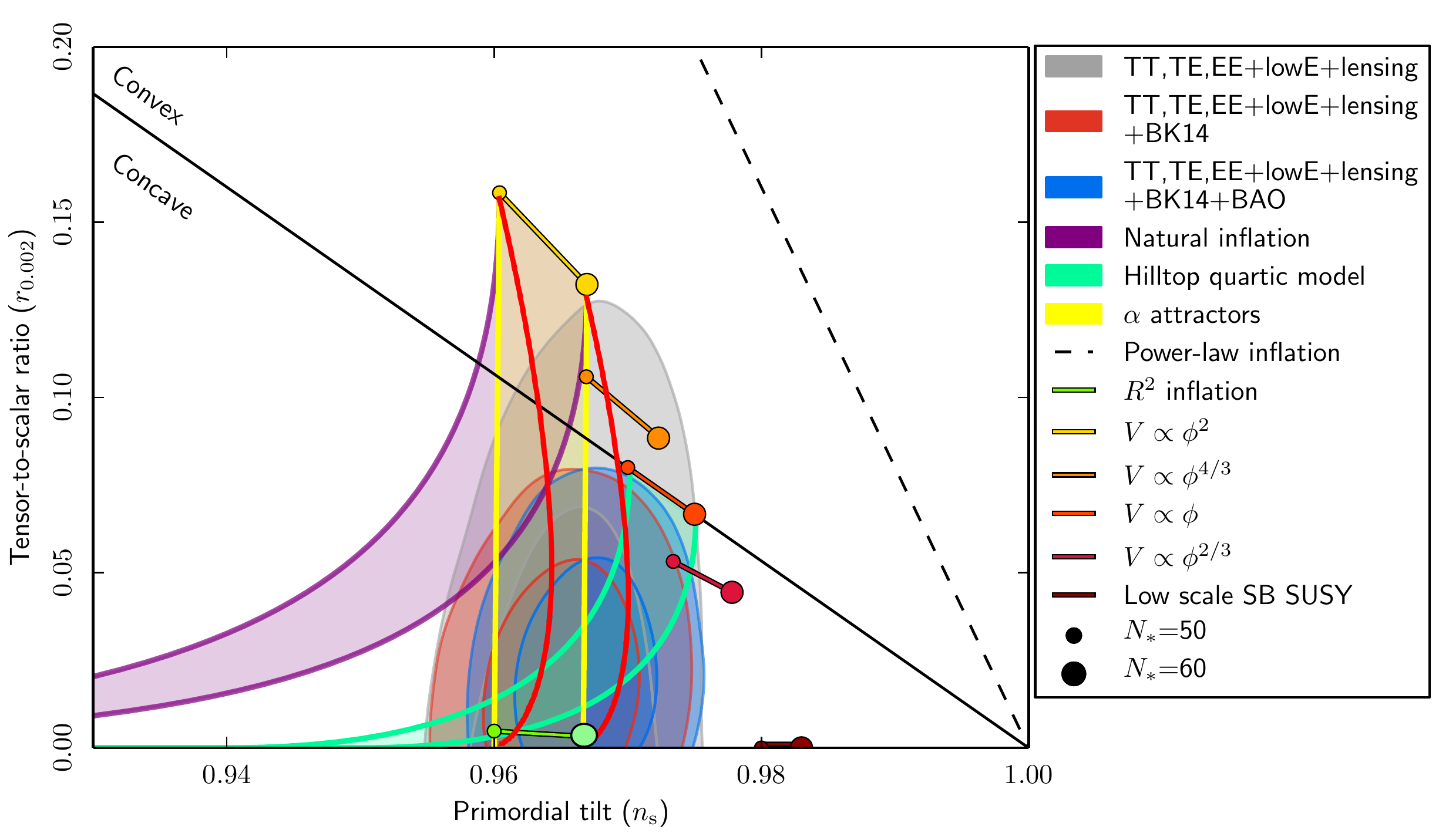}
\end{center}
\vskip -0.5cm 
\caption{\footnotesize Red lines show predictions of the simplest E-models \rf{E} for $N= 50$ and $N = 60$. These lines go very close to the yellow lines corresponding to the simplest T-model \rf{T}. These two basic $\alpha$-attractor models together cover a significant part of the area favored by Planck 2018.}
\label{F00}
\end{figure}

Note that the predictions of the E-models \rf{E} coincide with the predictions of the T-models \rf{T} in the limit $\alpha \to 0$ and $\alpha \to \infty$. In the intermediate range of $\alpha$, the simplest E-models \rf{E} predict slightly larger values of $n_{s}$ than the simplest T-models \rf{T}. These two models together cover a significant part of the area in the $n_{s},r$ space favored by Planck 2018, see Fig. \ref{F00}.

The predictions of these two classes of models nearly coincide with the predictions of the Starobinsky model and Higgs inflation model for $\alpha = 1$, but $\alpha$-attractors  allow much greater flexibility with respect to the tensor to scalar ratio $r$.  In this respect, $\alpha$-attractors are more ``future-safe'', allowing to   describe and parametrize various outcomes of the B-mode searches.

\section{Hilltop inflation after Planck 2018}

Hilltop inflation may seem  to provide a similarly good fit  to the Planck data.  The green band representing it is prominently shown in Fig. 8 of the Planck 2018 paper on inflation, which we reproduced  in Fig. \ref{F0} of our paper. It is also shown in Fig. 10 of the CMB-S4 Science Book \cite{Abazajian:2016yjj} and in Fig. 2.2 in PICO paper \cite{Hanany:2019lle}.
 
The first example of hilltop inflation models was given by the Coleman-Weinberg potential used in the new inflation scenario \cite{Linde:1981mu}
\be\label{CW}
V= {V_{0}}\,\left(1+ {\phi^{4}\over m^{4}}\bigl(2\log{\phi^{2}\over m^{2}} -1\bigr) \right) \ .
\ee
Later on, it became customary to consider  hilltop potentials of a more general type, 
\be\label{sloppy}
V= V_{0}\,\left(1-{\phi^{n}\over m^{n}}+...\right) \ ,
\ee
where the extra terms indicated by ... are supposed to be responsible for creating a minimum of  the potential \cite{Kinney:1995cc,Boubekeur:2005zm}. The simplest possibility is that such terms are higher order in $\phi^{n}\over m^{n}$.
For $n = 4$, this potential well represents the behavior of the Coleman-Weinberg potential near the top of the potential. One can show that for the small field models with $m \ll 1$, inflation occurs at $\phi \ll m$, where the higher order terms are negligible, and therefore some uncertainty in the definition of the potential at $\phi \sim m$ does not affect inflationary predictions.

That  is why the calculation of $n_{s}$ and $r$ in many papers on this issue, including the Planck 2018 paper on inflation  \cite{Akrami:2018odb}, is performed for the simplest models
\be\label{nodots}
V= V_{0}\,\left(1-{\phi^{n}\over m^{n}}\right) \ ,
\ee
ignoring the  terms indicated by $...$.
However, for $m \lesssim 1$, the most popular hilltop models with $n =  4$ shown in Fig. \ref{F0} predict $n_{s} = 0.94$ for $N = 50$ and $n_{s} = 0.95$ for $N=60$. Such models are ruled out by observational data. 

These predictions change for $m \gtrsim 1$,  but they approach safer values   $n_{s} \gtrsim 0.96$ favored by Planck 2018 only for $m \gtrsim 10$. In the large   $m$ limit the green lines describing predictions of this model in Fig. \ref{F0}  converge at the red circles corresponding to the predictions of inflation in the theory with a linear potential $V \sim \phi$. Moreover, a similar result is correct not only for $n=4$, but  for all hilltop potentials \rf{nodots}  \cite{Martin:2013tda}.   How can the complicated theories \rf{nodots} in the large $m$ limit give the same prediction  as the theory with a simple potential $V \sim \phi$? What is going on?

\begin{figure}[h!]
\begin{center}
\includegraphics[scale=0.56]{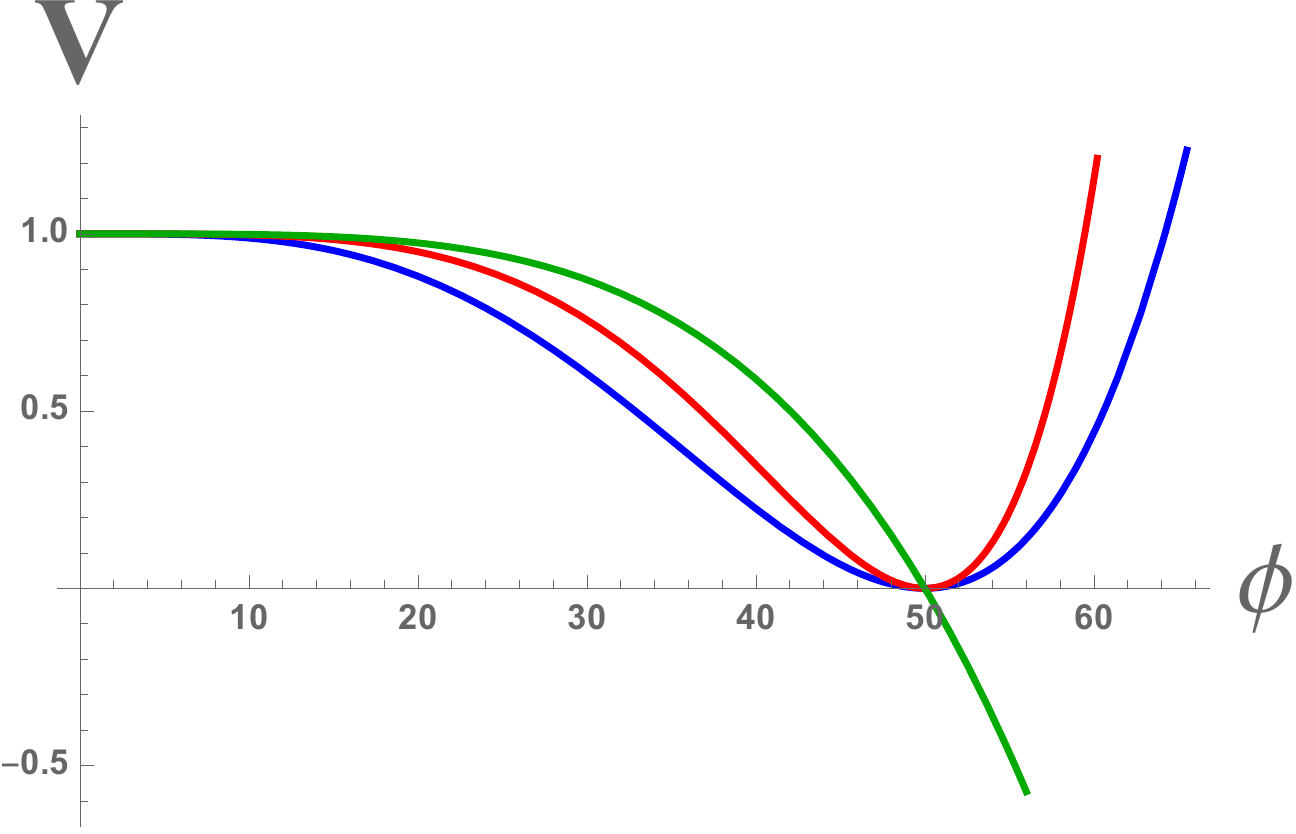}
\end{center}
\vskip -0.5cm 
\caption{\footnotesize Three hilltop potentials for $V_{0}= 1$ with $m = 50$. The  hilltop potential \rf{nodots} unbounded from below is shown by the   green line. The Coleman-Weinberg potential \rf{CW} is shown by the blue line, and the quartic Higgs-type version of the hilltop potential \rf{hill} is shown by the red line.}
\label{F9}
\end{figure}

To answer this question, let us look at the  the potential \rf{nodots}, which is shown by the green line in Fig. \ref{F9}. This potential   has a maximum (hilltop) at $\phi = 0$, and then $V(\phi)$ decreases and becomes zero at $\phi = m$. 
Because the potential does not have any minimum at $\phi \sim m$,  the potential at $\phi \approx m$ can be well approximated by a straight line. This approximation becomes better and better at large $m$, since the increase of $m$ stretches the potential horizontally.  For $m \gg 10$, the last 50 e-foldings in this scenario are effectively described by a linear potential proportional to $m-\phi$.  In this sense, the name ``hilltop inflation'' becomes a misnomer. The last 50 e-foldings in this scenario occur when the field moves down from $\phi \approx m -10$.  The slow-roll parameter $\epsilon$ in this effective theory is given by $ {1\over 2(m-\phi)^{2}}$, it is smaller than 1 and  inflation continues until the point $m-\phi_{\rm end} \approx 1/\sqrt 2$.  Investigation of inflation in this scenario could suggest that its predictions provide a good match to Planck data. But this conclusion would be premature because such models suffer from the graceful exit problem. 

Indeed, after the end of inflation at $m-\phi_{\rm end} \approx 1/\sqrt 2$ the field experiences free fall, reaches  $V(\phi) = 0$  and then continues falling all the way down. During this process,  the universe starts contracting, and it collapses within the time of the order of the Hubble time $H^{-1}$ at the end of inflation \cite{Felder:2002jk}, producing no CMB and no observers. Thus there is a direct relation between the remarkable universality of the inflationary predictions of the models of this class and their fundamental inconsistency. 

This result is rather general. It applies, in the large $m$ limit, to {\it all} potentials $V(\phi/m)$ where inflation occurs along the slope of the potential unbounded from below.  Such models do not lead to acceptable inflationary cosmology, so they should be excluded from consideration.

To  achieve graceful exit in a consistent theory of the hilltop inflation,  one may return to the Coleman-Weinberg potential \rf{CW}, or consider an improved hilltop potential
\be\label{hill}
V= V_{0}\,\left(1-{\phi^{n}\over m^{n}}\right)^{2} \ ,
\ee
which is  given by $V_{0}(1-2{\phi^{n}\over m^{n}}+...)$ at small $\phi$. All three hilltop potentials \rf{CW}, \rf{nodots} and \rf{hill} vanish at $\phi = m$, but their behavior in the vicinity of the point $\phi = m$ is dramatically different. The potential \rf{nodots} is shown in Fig. \ref{F9} by a green line, to match the color of predictions of this model in the Planck 2018 figure Fig. \ref{F0}. The Coleman-Weinberg potential \rf{CW}  
is shown by the blue line, and the regularized hilltop potential \rf{hill} is shown by the red line.

As we see, the non-regularized potential \rf{nodots} goes straight down at $\phi = m$ where $V= 0$; it is approximately linear at $\phi \sim m$. Meanwhile any natural minimum of the potential is approximately quadratic, with the curvature proportional to $m^{2}$, see Fig.~\ref{F9}. Therefore one may expect that in the large $m$ limit, the predictions of the stabilized hilltop  inflation models with a quadratic minimum should converge to predictions of the chaotic inflation potential {\it with a quadratic potential}, shown by the yellow circles in the Planck 2018 figure Fig.~\ref{F0}.  But these predictions are 2 times higher than the predictions of the  linear potential! This conclusion is quite general; it applies to all single parameter models $V({\phi\over m})$.  
\begin{figure}[h!]
\begin{center}
\includegraphics[scale=0.4]{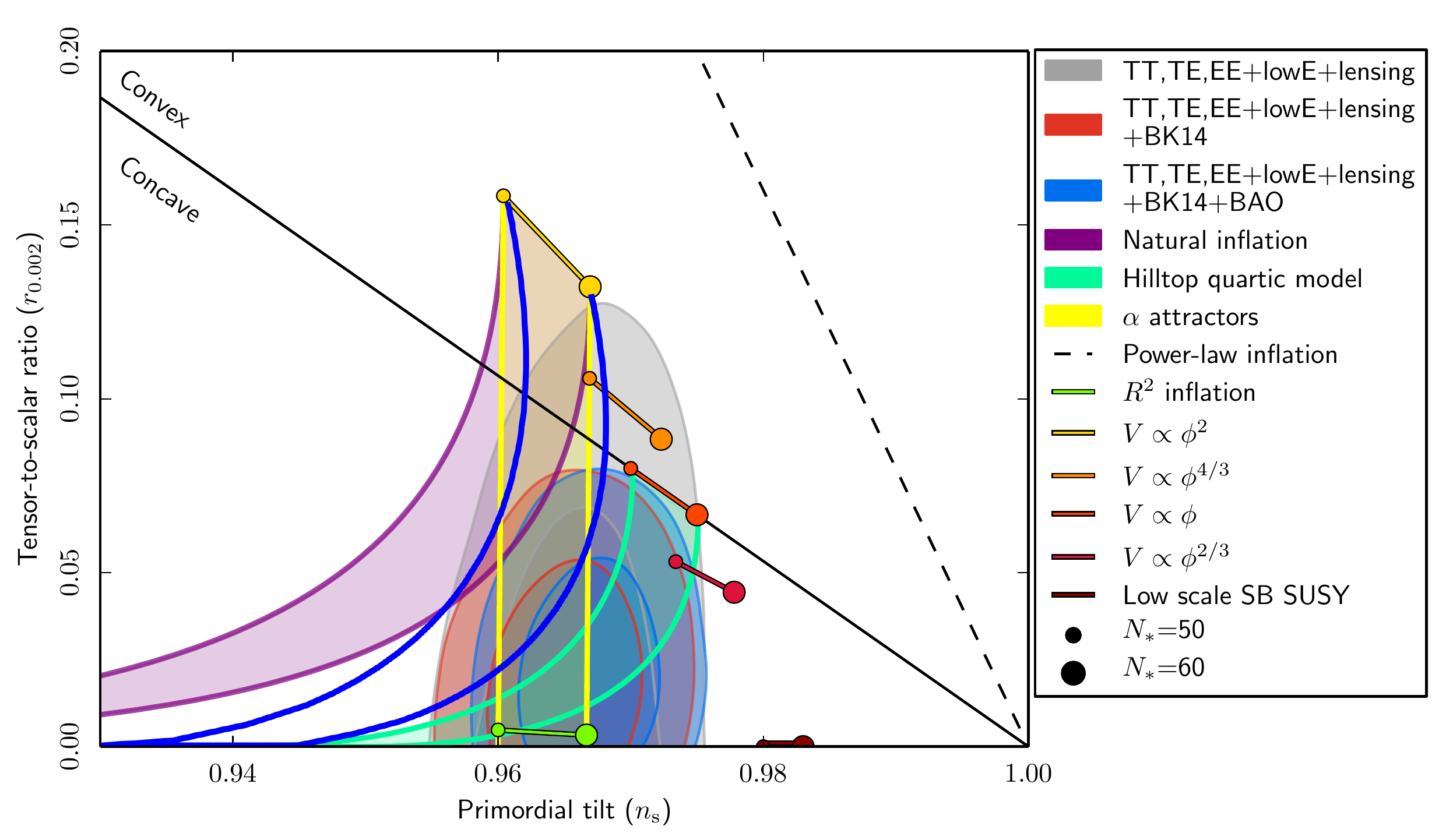}
\end{center}
\vskip -0.5cm 
\caption{\footnotesize Thick blue lines falling down from the two yellow circles correspond to $n_{s}$ and $r$ for the Coleman-Weinberg potential \rf{CW} for $N= 50$ (left line) and $N = 60$ (right line). }
\label{F10}
\end{figure}

To check these general expectations,  we   begin with evaluation of $n_{s}$ and $r$ for the Coleman-Weinberg potential \rf{CW}. The results of our analysis are shown by two thick blue lines in  Fig. \ref{F10}, for $N= 50$ (left line) and $N = 60$ (right line). As one can see, the area between these lines only minimally overlaps with the 1$\sigma$ region favored by the Planck data. Thus this model is considerably disfavored as compared to the (unphysical) hilltop model shown by the green area in this figure. 

 \begin{figure}[h!]
\begin{center}
\includegraphics[scale=0.4]{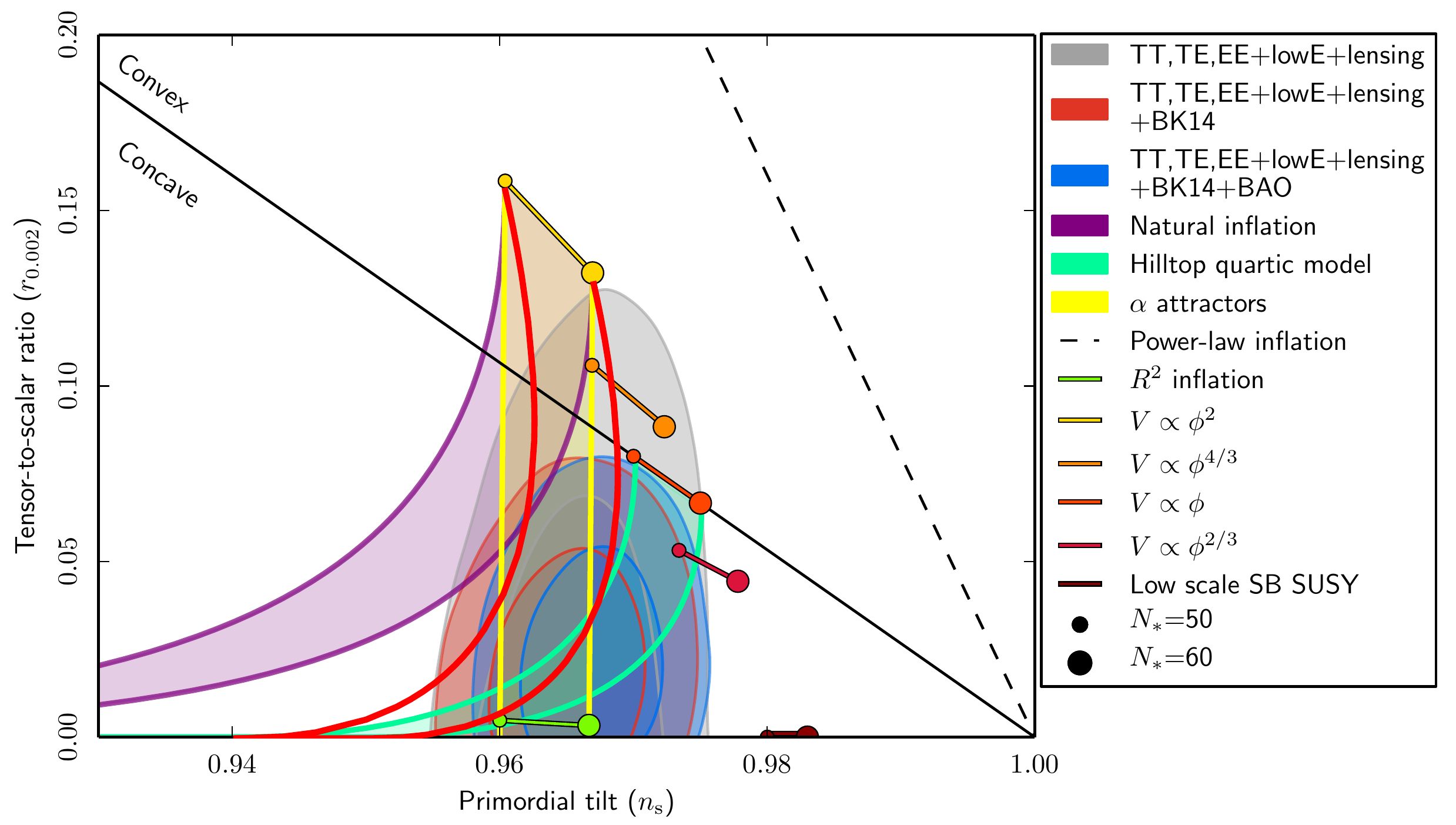}
\end{center}
\vskip -0.5cm 
\caption{\footnotesize Thick red lines falling down from the two yellow circles correspond to $n_{s}$ and $r$ for the regularized hilltop potential \rf{hill} for $N= 50$ (left line) and $N = 60$ (right line).}
\label{F11}
\end{figure}

Fig. \ref{F11} describes  $n_{s}$ and $r$ for the hilltop potential \rf{hill} for $n=4$.  The thick red lines correspond to $n_{s}$ and $r$ for $N= 50$ (the left line) and $N = 60$ (the right line). As one can see, the predictions of this model for $n_{s} >  0.94$ are quite different from the predictions of the unphysical model \rf{nodots}, but it provides a  better fit to the Planck 2018 data than the Coleman-Weinberg model \rf{CW}.
However, whereas one can find some physical motivation for the Coleman-Weinberg model (scale invariance broken by quantum corrections), the model \rf{hill} looks rather {\it ad hoc}, being specifically designed to mitigate the problems of the hilltop models  \rf{nodots}. 

One can avoid this conclusion by introducing additional mass parameters, or by considerably modifying the general structure of the potential, but in this paper we are trying to concentrate on relatively simple models which do not require additional parameters, twists and turns for explanation of the observational data.  

\begin{figure}[h!]
\begin{center}
\includegraphics[scale=0.5]{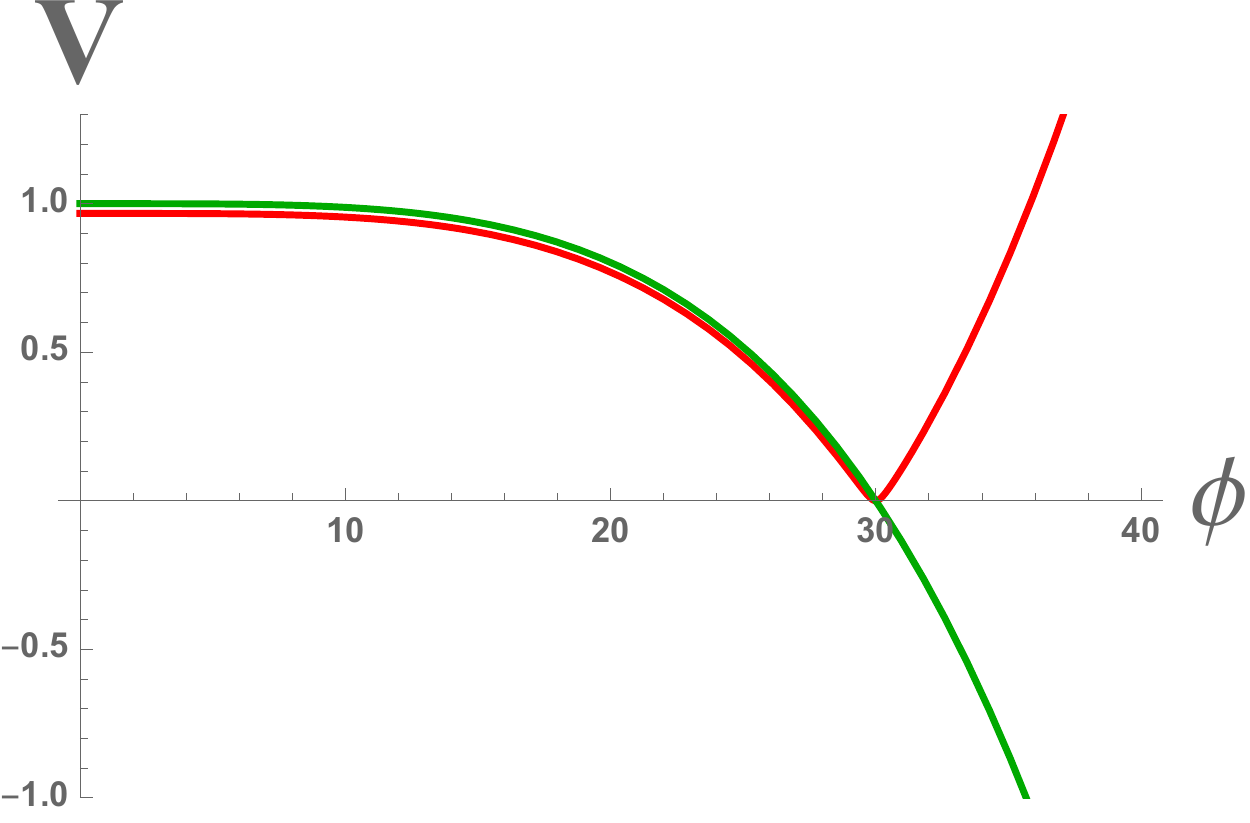}
\end{center}
\vskip -0.5cm 
\caption{\footnotesize The  hilltop potential $V_{0} (1-{\phi^4\over m^4})$  \rf{nodots} unbounded from below  is shown by the   green line  for  $V_{0} = 1$. The strongly modified hilltop potential \rf{sqrt} is shown by the  red line for  $m = 30$, $c = m^{{-1}}$. }
\label{Faa}
\end{figure}

Nothing can explain this point better than an explicit example of a theory which can preserve the predictions of the naive hilltop model \rf{nodots} while providing vacuum stabilization at $V = 0$. To do it, one should keep the potential $1-{\phi^{4}\over m^{4}}$ intact all the way down to a small vicinity of $\phi = m$, and then force the potential to turn up very sharply. Here is the simplest potential satisfying this property that we were able to invent: 
\be\label{sqrt}
V_{\rm mod}= V_{0}\, \Bigl(\sqrt{ { c^{2}}+ \Bigl(1-{\phi^{4}\over m^{4}}\Bigr)^{2}} -{c}\Bigr) .
\ee
Here $c$ is a small positive constant, which is introduced to make the potential smooth at its minimum at $\phi = m$.  The potential $V_{\rm mod}$ is everywhere positive. In the limit $c \ll m^{{-1}}$, this potential \rf{sqrt}  is equal to the absolute value of the potential \rf{nodots}, see Fig.~\ref{Faa}.

Predictions of this theory for $c \lesssim m^{{-1}}$ are well described by the green area in Fig. \ref{F0}. 
This potential  contains the potential \rf{hill} under the square root. But if, as we admitted, the potential \rf{hill} is not well motivated, then it is  hard to    say anything positive about the potential \rf{sqrt}.  It is constructed by a set of additional manipulations with the potential  \rf{hill}, with the sole purpose to save the predictions of the physically inconsistent models \rf{nodots}. But why would anybody want to do it?

One can summarize these considerations as follows. The reason why the original hilltop models \rf{sloppy} with $n = 4$ and $m \lesssim 1$ were popular and deserved to be a target for future searches was that they were ultimately simple. One could modify the shape of the potential at $\phi \sim m$ without modifying its inflationary predictions, and the value of $r= 8 {m^{4} \over N^3}$ could be easily changed by varying $m$  without affecting the attractor value $n_{s} = 1-3/N$. This value of $n_{s}$ was consistent with WMAP.

However, the spectral index  $n_{s} = 1-3/N$ was only marginally consistent with Planck 2013, and  it was finally ruled out by Planck 2015 and Planck 2018. In this sense,  the original beauty of hilltop models  is long gone.  The simplest attempts to cure the problems of the hilltop models \rf{nodots} with $n = 4$ lead  to a family of models with the predictions interpolating between two bad attractor points. In the limit $m \ll 1$ these models predict unacceptably small $n_{s}$, whereas in the large $m$ limit they predict unacceptably large $r = 8/N$,   two times greater than the predictions of the simplest (but inconsistent) hilltop model  \rf{nodots}. 

One can design the models providing a good fit to the data, such as \rf{hill}, but the predictions of such models are mostly determined not by the structure of the potential at the hilltop, but by the previously unspecified model-dependent terms shown by ... in \rf{sloppy}. And their predictions are very different from the predictions of the original hilltop inflation models \rf{nodots} shown by the green area in Fig. \ref{F0}. One can further modify these models, by requiring that their predictions should be  consistent with the predictions of the models \rf{nodots} with a potential unbounded from below, but this additional  requirement is  unmotivated. Therefore we would argue  the green area in  Fig. \ref{F0} does not properly represent predictions of the simplest consistent hilltop models discussed in  \cite{Akrami:2018odb}.

And this is not the only problem of the hilltop models.  Inflationary predictions described here where evaluated under the assumption that inflation begins at the hilltop. Whereas it is possible for inflation to begin  at a local maximum of a potential even if its height is 10 orders of magnitude below the Planck density \cite{Linde:2004nz,Linde:2014nna,East:2015ggf,Linde:2017pwt}, it is arguably much simpler for inflation to begin at the nearly Planckian density, as in the simplest versions of chaotic inflation \cite{Linde:1983gd,Linde:2005ht}. This may happen in the models \rf{CW} and  \rf{hill} if inflation begins at $\phi \gg m$. But in that case inflation ends at $\phi > m$, and the field never rolls up to the hilltop at $\phi = 0$ to start a new stage of inflation there. Therefore inflationary predictions become similar to the predictions of the chaotic inflation models $\phi^{n}$ with $n>2$. Such models are already ruled out.

Thus to make the hilltop models attractive one should modify them even further, in order to prevent inflation starting at $\phi \gg m$. There are many ways to do it, for example by making the potentials exponentially steep at $\phi \gg m$, but this would make consistent models of hilltop inflation even more contrived. 

\section{D-brane inflation}

The string theory origin of D-brane inflation model is often attributed to the KKLMMT model~\cite{Kachru:2003sx}, where D3-brane - $\overline {\rm D3}$-brane interaction was studied in the context of the volume modulus stabilization. Earlier proposals for D-brane inflation relevant to our current discussion  were made in~\cite{Dvali:2001fw,Burgess:2001fx,GarciaBellido:2001ky}. Whereas string theory interpretation of D-brane inflation is somewhat unsettled, the positive evaluation  of the cosmological consequences of these models in  \cite{Akrami:2018odb} prompted us to revisit them and implement these models in supergravity   \cite{Kallosh:2018zsi}.

 We will discuss two classes of these models, using terminology established in  \cite {Martin:2013tda}. The first class of models called BI (for brane inflation) has the Coulomb-type potential 
 \be\label{BI}
 V_{{\rm BI}} = V_{0}\left(1-{m^{n}\over \phi^{{n}}}\right) \ .
 \ee
BI models has a potential unbounded from below. In the large $m$ limit   predictions of these models coincide with the predictions of the theory with a linear potential, but BI models suffer  from the same  problem as the hilltop inflation \rf{nodots}, and therefore they require  a consistent generalization. 

 This generalization was proposed in the KKLMMT paper \cite{Kachru:2003sx}   where the potential takes a form of the inverse harmonic function; in  \cite {Martin:2013tda} it was called KKLTI (for KKLT inflation):
 \be
V_{KKLTI} =   V_{0}\left(1+{m^{n}\over \phi^{{n}}}\right)^{{-1}} = V_{0}\, {\phi^n\over   \phi^n+ m^n}  \ .
\label{KKLTI}
\ee
Similar potentials may appear in a different context as well, see e.g. \cite{Dong:2010in} and Appendix A in \cite{Kallosh:2018zsi}.

\begin{figure}[!h]
\hspace{-3mm}
\begin{center}
\includegraphics[width=9cm]{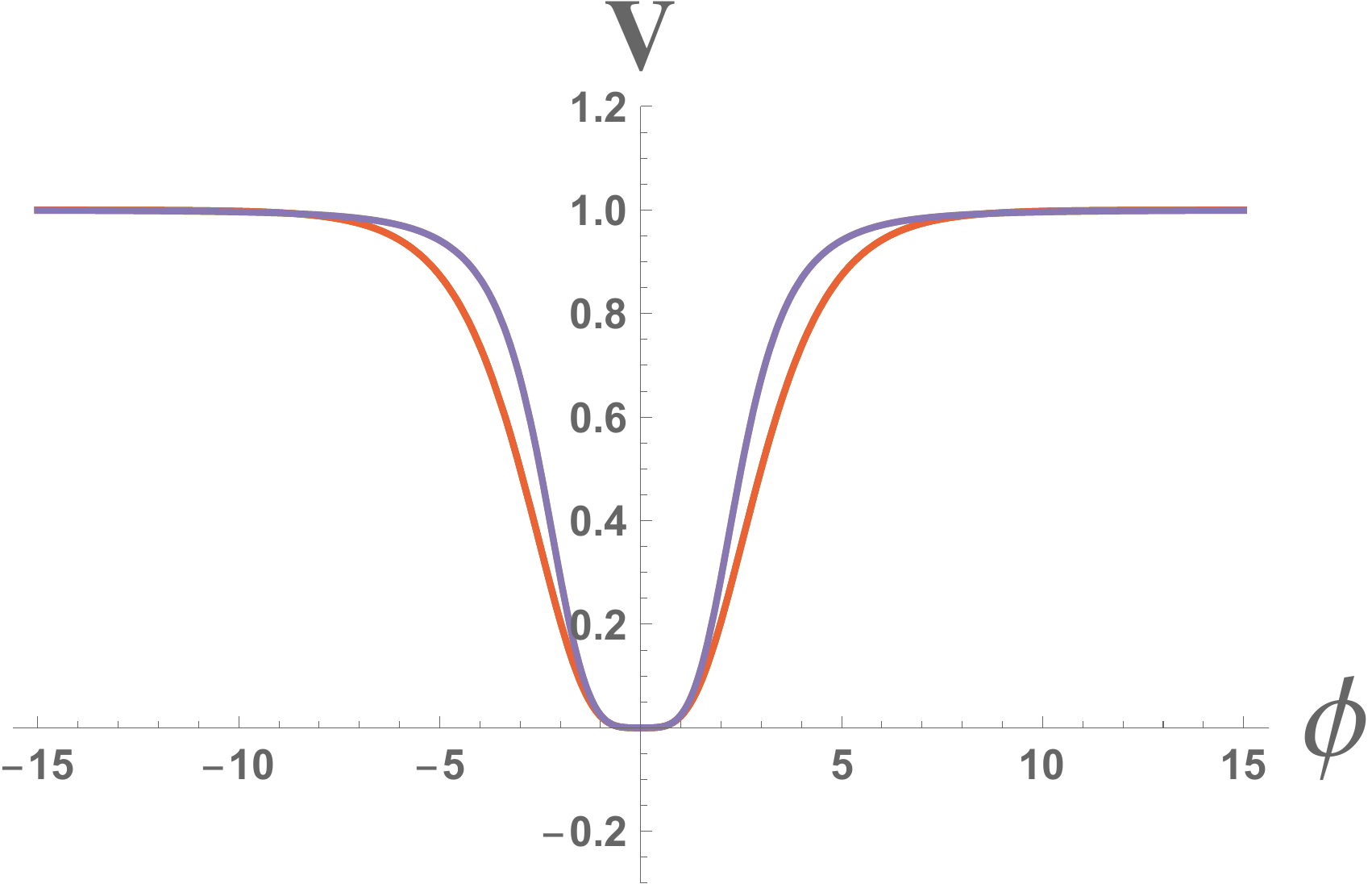}
\caption{\footnotesize  The red line shows the quartic $\alpha$-attractor potential $V= \tanh^4  {\phi \over \sqrt{6 \alpha} } $ with $\alpha =1$.  The dark blue line shows  the KKLTI potential $V= {\phi^4\over \phi^4 + m^{4}}$ for $m=2.5$.  }
\label{Compare}
\end{center}
\vspace{0cm}
\end{figure}

It is instructive to compare KKLTI models with $\alpha$-attractors, see Fig.~\ref{Compare}. It shows that  $\alpha$-attractors and D-brane inflation models have plateau potentials which look very similar, but the KKLTI   models have potentials approaching the plateau polynomially rather than exponentially.

In the small $m$ limit, D-brane models behave as cosmological attractors with  the attractor point at
$ n_s \approx 1- {2(n+1) \over N (n+2)}$.
For odd values of $n$, both BI and KKLTI models have potentials unbounded from below at $\phi < 0$ and therefore require further modifications; we will not discuss them here.   The most interesting of the  KKLTI models is the one    with $n= 4$, though the model with $n = 2$ also deserves  attention. For $n=4$ one has at attractor point
\be\label{nsrkklt4}
 n_s = 1- {5\over3 N}, \qquad r = {4 m^{4/3}\over (3N)^{4/3}} \ .
\ee
For $n = 2$, the attractor is at 
\be\label{nsrkklt2}
 n_s = 1- {3\over2 N}, \qquad r = {\sqrt 2 m\over N ^{3/2}} \ .
\ee

In the large $m$ limit, the predictions of BI and KKLTI models are very different. Indeed, predictions of the BI models in the large $m$ limit converge to the predictions of the models with $V\sim \phi$. This is as misleading as the similar predictions of the hilltop inflation models \rf{nodots}, which yield $r= 4/N$.  In the KKLTI models in the large $m$ limit the last 50-60 e-folds of inflation occur at  $\phi = O(10)$. Therefore for $m \gg 10$ the last 50-60  e-foldings occur for $1\ll \phi\ll m$, where the KKLTI potential is proportional to $\phi^{n}$. Therefore for $n = 2$ in the large $m$ limit of the KKLTI models one has $r = 8/N$, and for $n = 4$ one has $r = 8/N$, which is 4 times greater than the value of $r$ in the large $m$ limit for the unphysical   BI models \rf{BI}. 
 \begin{figure}[h!]
\begin{center}
\includegraphics[scale=0.4]{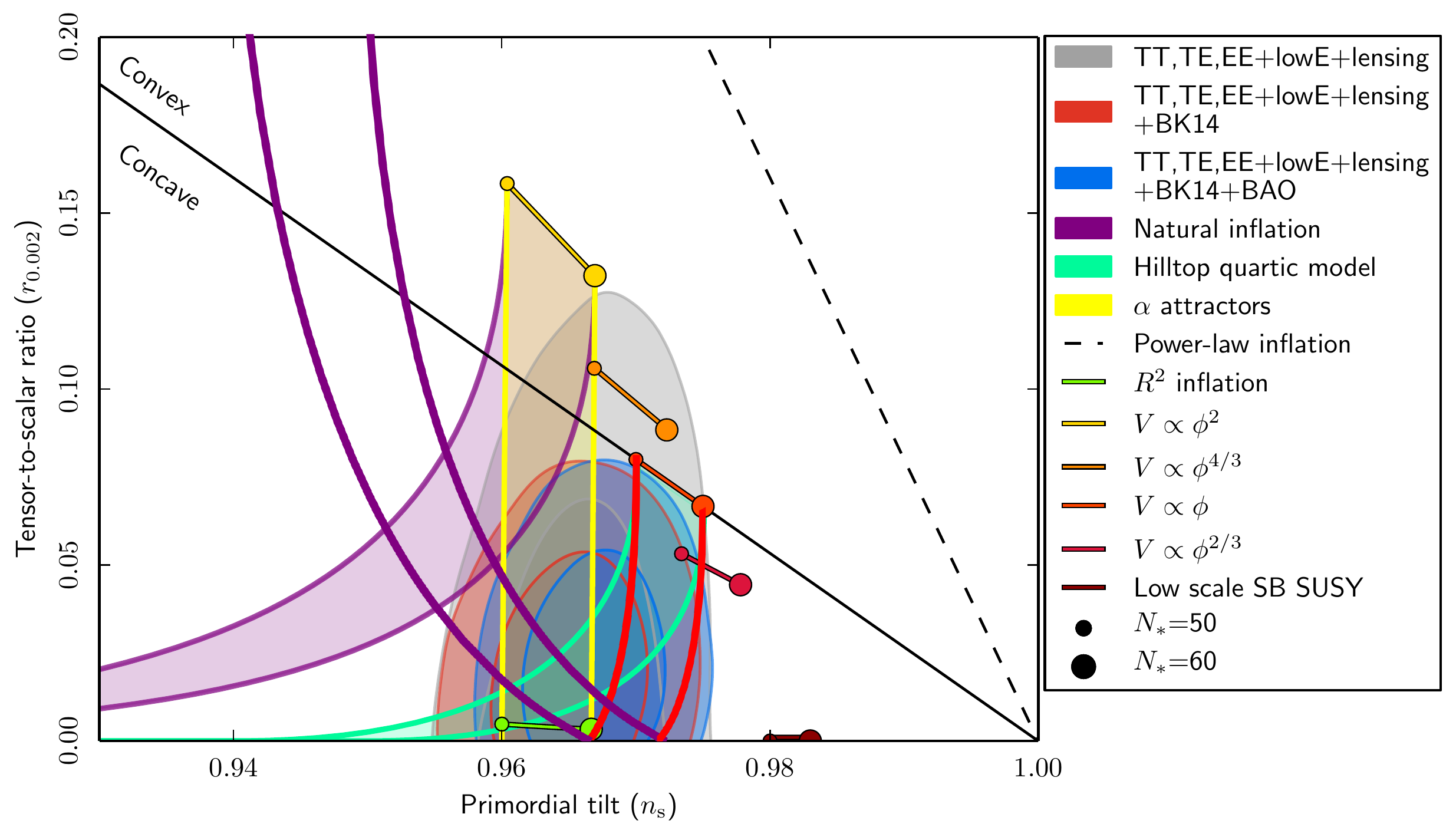}
\end{center}
\vskip -0.3cm 
\caption{\footnotesize Red lines correspond to predictions of the BI theory for $n = 4$, $N = 50$ and $N = 60$. This  theory is inconsistent and requires modification provided by the KKLTI model. Predictions of the KKLTI model are shown by the thick purple lines converging with the red lines at the attractor points $r = 0$, $n_{s} =  1- {5 \over 3 N}$. }
\label{F12}
\end{figure}

To illustrate this difference, we show the predictions of the BI model and the KKLTI model for $n=4$ in Fig. \ref{F12}. As one can see, observational predictions of the BI model do not  have much in common with the predictions of consistent D-brane models of the type of KKLTI  for $r > 10^{{-2}}$. The predictions of these models converge at $r \lesssim 10^{{-3}}$  \cite{Kallosh:2018zsi}.  Since this difference of the predictions of KKLTI and BI models  is rooted in the physical inconsistency of the BI models,  we will no longer investigate BI models in this paper. 

\begin{figure}[!h] 
\begin{center}
\includegraphics[scale=0.32]{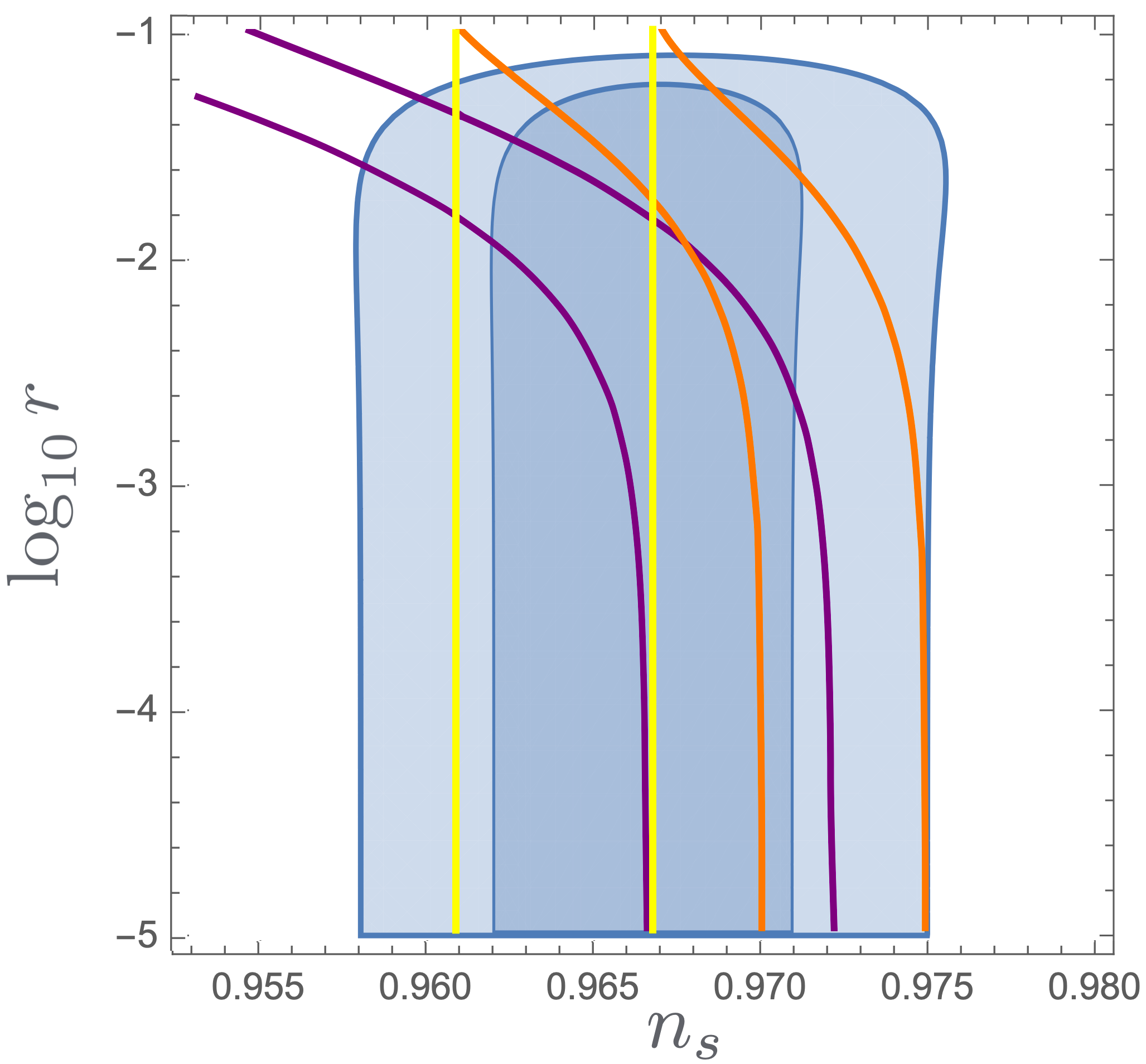} \hskip 30pt
\includegraphics[scale=0.32]{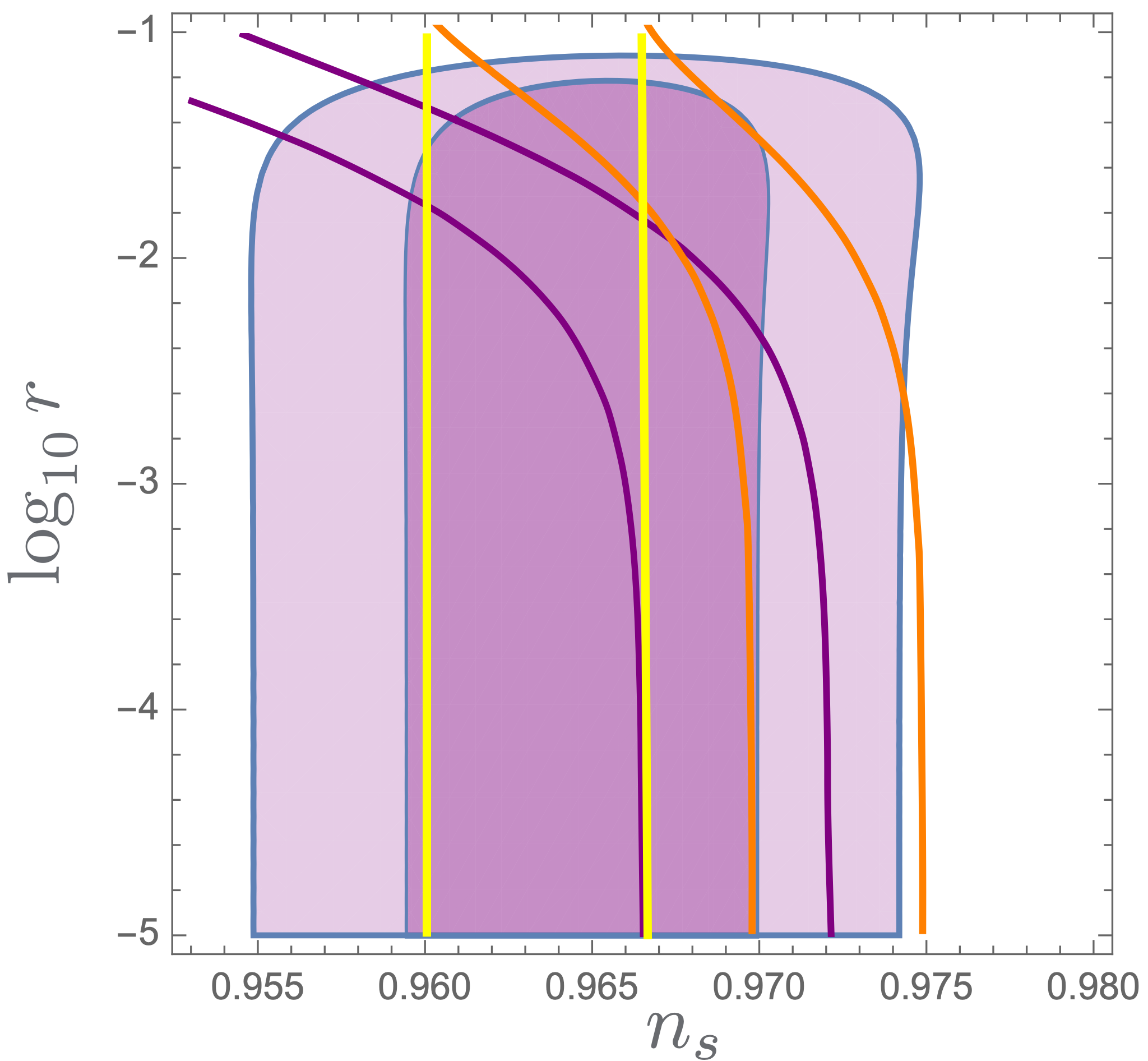}
\end{center}
 \vspace{-0.3cm}
\caption{\footnotesize Comparison of predictions of $\alpha$-attractors  and  of the D-brane inflationary models within the $2\sigma$ area of  the Planck 2018 results for $n_{s}$ and $r$. On the left panel, the dark (light) blue area is the Planck 2018 $1\sigma$  ($2\sigma$) region, with an account taken of the CMB-related data and baryon acoustic oscillations (BAO).  The right panel represents the Planck 2018 results based on the CMB related data only, without  BAO. These data, without BAO,  were used by Planck 2018 in  \cite{Akrami:2018odb}  to evaluate inflationary models. Two yellow lines on both panels are for the quadratic T-model of $\alpha$-attractors for $N  = 50$ and $N  = 60$. Two purple lines are for the $n=4$ KKLTI model. Two orange lines  show the predictions of the $n=2$ KKLTI model.  }
\label{Quartic}
\end{figure}

It is very instructive to study KKLTI models in parallel with $\alpha$-attractors by plotting their predictions
not   for $r$ but for $\log_{10} r$, all the way down to $r \sim 10^{{-5}}$, see Fig. \ref{Quartic} taken from our paper with Yusuke Yamada  \cite{Kallosh:2018zsi}. 

As one can see  from Fig. \ref{Quartic}, $\alpha$-attractors and KKLTI models can describe arbitrarily small values of $r$.  Evaluation  of $n_{s}$ for these models in the range of $N$ from 50 to 60 gives us an unexpected numerological bonus: The $\alpha$-attractor prediction  $n_{s} = 1-{2\over N}$ for $N = 60$ exactly coincides with the prediction  $ n_s = 1- {5\over3 N}$ of the $n = 4$ KKLTI model  for $N = 50$. Therefore $\alpha$-attractors and KKLTI with $n = 4$ stay ``shoulder to shoulder'', covering most of the  $(n_{s}, r)$ space favored by Planck. They miss only a small upper right part of the range of the data favored by Planck 2018, but the KKLTI model with $n = 2$ completes this job. As a result,  $\alpha$-attractors, in combination with the KKLTI models,  almost completely cover the dark blue (red) $1\sigma$ region favored by Planck~2018.

\section{Discussion}

In this paper we made an attempt to examine the models favored by the latest Planck 2018 data release, including $\alpha$-attractors, hilltop models and D-brane inflation models. Since $\alpha$-attractors were studied in many of our previous papers, we limited ourselves to a brief review of the simplest T-models and E-models in their relation to the Planck 2018 data. We found that a combination of these  two basic $\alpha$-attractor models  covers a significant part of the area in the ($n_{s}, r$) space favored by Planck 2018, see Fig. \ref{F00}.

The results of our analysis of hilltop models and D-brain models are somewhat unexpected. We were puzzled by the fact that the predictions of the  hilltop models $V = V_{0}(1-{\phi^{n}\over m^{{n}}})$,  prominently represented by the green area in Fig. \ref{F0} (for $n = 4$), as well as the predictions of the  D-brane inflation models with $V = V_{0}(1-{m^{n}\over \phi^{{n}}})$, in the large $m$ limit converge to the predictions of the simplest  inflationary model with a linear potential $V \sim \phi$, as is shown by the thick green lines for the hilltop models in the   Fig. \ref{F0} and by the thick red lines in \ref{F12}  for the D-brane models. 
However,  we found that this result is a direct consequence of the unboundedness of these potentials from below.

One can  avoid this inconsistency by bending these potentials and creating a minimum at $V = 0$.  For $m \ll 1$, inflation occurs at $\phi \ll m \ll 1$, so one can modify these models at $\phi \sim m$ without modifying their inflationary predictions. However, the typical hilltop models with $m \lesssim 1$ are ruled out by observations. Meanwhile we found that the improved hilltop models with  a minimum at $V = 0$ in the large $m$ limit typically predict $r= 8/N$, which is way too large. Since the predictions of the simplest consistent versions of the hilltop inflation  are considerably different from the predictions of the simplest (but inconsistent) hilltop inflation model $V = V_{0}(1-{\phi^{4}\over m^{{4}}})$ shown by the green area in Fig. \ref{F0}, 
we would argue that  one should no longer associate  hilltop inflation with the green area shown in  Fig. \ref{F0}  in Planck 2018  \cite{Akrami:2018odb}, as well as in the related figures shown by CMB-S4 and PICO \cite{Abazajian:2016yjj,Abazajian:2019eic,Hanany:2019lle}.\footnote{Interestingly,  in the  first version of the Planck 2013 paper on inflation, which can be found  \href{https://arxiv.org/pdf/1303.5082v1.pdf}{here}, Fig. 1 did {\it not}  contain the green area. It was added only later, upon the referee request.}

Similarly, the  simplest  versions of  the D-brain inflation with the potential $V = V_{0}(1-{m^{n}\over \phi^{{n}}})$ are physically inconsistent. Consistent generalizations of this model with the positively definite potential \cite{Kachru:2003sx} in the large $m$ predict unacceptably large $r = 4n/N$; see a recent discussion of these models, called KKLTI models, in \cite{Kallosh:2018zsi}. However,  at $m \lesssim 1$ the predictions of the KKLTI models  converge to  $ n_s \approx 1- {2(n+1) \over N(n+2) }$, which provides a very good fit to the Planck data for $n = 4$, and an acceptably  good fit for $n = 2$.  As one can see from Fig. \ref{Quartic}, the simplest  $\alpha$-attractors models, in combination with the KKLTI models,  almost completely cover the $1\sigma$ region favored by Planck~2018  \cite{Kallosh:2018zsi}.

The recent Astro2020 Science White Paper ``Probing the origin of our Universe through cosmic microwave background constraints on gravitational waves'' \cite{Shandera:2019ufi} suggested that most of the textbook inflation models cannot describe inflation with $r \lesssim 10^{{-3}}$. As one can see from Fig. \ref{Quartic}, this problem does not appear in the broad class of $\alpha$-attractor models, as well as in the KKLTI models discussed above. Therefore these two classes of models can be very useful for parametrizing and describing the results of the future search of the gravitational waves produced by inflation, independently of their outcome. We will return to the discussion of this issue in a forthcoming publication.
\vskip 0.3cm

{\bf {Acknowledgments:}} We are grateful to  Z. Ahmed, D. Green, F. Finelli, R. Flauger, M. Hazumi, L. Knox, Chao-Lin Kuo, J. Martin, L. Page,  L. Senatore,  E. Silverstein, V. Vennin,   and R. Wechsler   for stimulating discussions, and to J.~J.~Carrasco, S. Ferrara,  D. Roest, T. Wrase and Y. Yamada for collaboration on related work.  This work is supported by SITP,  by the US National Science Foundation grant PHY-1720397 and by the Simons Foundation Origins of the Universe program (Modern Inflationary Cosmology collaboration).
 
\parskip 3pt 
   
\bibliographystyle{JHEP}
\bibliography{lindekalloshrefs}

\end{document}